\documentclass[floats,floatfix,aps,pre,preprint, amsmath,amssymb]{revtex4}
\usepackage{graphicx}
\usepackage{dcolumn}
\usepackage{epstopdf}
\usepackage{bm}
\usepackage{graphicx}
\usepackage{subfigure}
\usepackage{color}
\usepackage{latexsym}
\usepackage{amsfonts}
\usepackage{amsmath}

\usepackage{amssymb,amsfonts,amsmath}

\def \>{\rangle} 
\def \<{\langle} 
 
\def\be{\begin{equation}} 
\def\ee{\end{equation}} 
\newcommand \bea {\begin{eqnarray} } 
\newcommand \eea {\end{eqnarray}}

\begin{document}

\title{ Information processing and signal integration in bacterial quorum sensing}
\author{Pankaj Mehta, Ned S. Wingreen}
\affiliation{Dept. of Molecular Biology, Princeton University, Princeton, NJ 08544}
\author{ Bonnie L. Bassler}
\affiliation{Dept. of Molecular Biology, Princeton University, Princeton, NJ 08544\\
Howard Hughes Medical Institute, Chevy Chase, Maryland, United States of America}

\author{Sidhartha Goyal, Tao Long}
\affiliation{Dept. of Physics, Princeton University, Princeton NJ, 08544}

\begin{abstract}

Bacteria communicate using secreted chemical signaling molecules called autoinducers in a process known as quorum sensing. The quorum-sensing network of the marine bacterium {\it Vibrio harveyi} employs three autoinducers, each known to encode distinct ecological information. Yet how cells integrate and interpret the information contained within the three autoinducer signals remains  a mystery.  Here, we develop a new framework for analyzing signal integration based on Information Theory and use it to analyze quorum sensing in {\it V. harveyi}. We quantify how much the cells can learn about individual autoinducers and explain the experimentally observed input-output relation of the {\it V. harveyi} quorum-sensing circuit. Our results suggest that  the need to limit interference between input signals places strong constraints on the architecture of bacterial signal-integration networks, and that bacteria likely have evolved active strategies for minimizing this interference. Here we analyze two such strategies: manipulation of autoinducer production and feedback on receptor number ratios.

\end{abstract}
\maketitle

\section{ Introduction}

Unicellular organisms live in complex and dynamic environments. They sense and respond to both external environmental cues and to each other through  quorum sensing, i.e. cell-to-cell communication. Adapting to changing environments often requires cells to simultaneously integrate information from multiple environmental inputs, and cells have developed elaborate signaling networks to accomplish this feat. However, the design principles underlying the architectures of these networks remain largely mysterious. For example, in the model quorum-sensing bacerium {\it Vibrio harveyi}, three chemical communication signals are integrated to regulate gene expression, but the logic and mechanism underlying this integration are poorly understood. Such open questions highlight the need for new conceptual and theoretical tools to supplement ongoing experimental work. Here, we present a new theoretical framework for understanding signal integration based on Information Theory  (Shannon, 1948) and we use it to study information processing in the {\it V. harveyi} quorum-sensing circuit.
 
Quorum sensing is widespread in the bacterial world and can occur both within and between bacterial species, and even between bacteria and their eukaryotic hosts (Waters and Bassler, 2005). Quorum sensing enables bacteria to alter their behavior depending on the number and/or species of bacteria present and is important for a variety of collective behaviors such as biofilm formation, bioluminescence,  virulence, and stress-responses (Waters and Bassler, 2005;   Bassler and Losick, 2006; Waters and Bassler, 2006). The {\it V. harveyi}  quorum-sensing circuit is among the best characterized of all quorum-sensing networks (Fig. \ref{fig:fig0}a). {\it V. harveyi}  produces and detects three chemical signaling molecules called autoinducers (AIs), AI-1, CAI-1, and AI-2. Whereas AI-1 is  produced only by {\it V. harveyi},  CAI-1 is produced by other  {\it Vibrios}, and AI-2 is produced by a large variety of both Gram-negative and Gram-positive bacteria and likely acts as a universal signaling molecule. Thus, the use of multiple AIs potentially provides bacteria with information about the local density of {\it V. harveyi},  all {\it Vibrios}, and total bacteria (Waters and Bassler, 2005). Sensory information from the three AIs is channeled through a common phosphorelay (see Fig. 1). The three autoinducers, AI-1, CAI-1, and CAI-1,  are detected by cognate transmembrane receptors, LuxN, LuxPQ, and CqsS, respectively (Henke and Bassler, 2004), and they collectively control production of the master quorum-sensing transcriptional regulator LuxR (see Fig. \ref{fig:fig0}a) (Tu and Bassler, 2007). 

Because all information about the AIs is  channeled through a common phosphorelay, it is unclear how much bacteria can learn about each individual input.  Even less clear is how the architecture and kinetic parameters (e.g., kinase and phosphatase rates) of the quorum-sensing network affect its signal-transduction properties. To address these questions, we have developed a new mathematical framework for analyzing signal integration in cells based on Information Theory (Shannon, 1946; MacKay, 2003). Information theory provides a natural language for formulating questions about information processing and signal integration. It is used extensively in engineering to model signaling in man-made communication devices and  has also proven to be a powerful tool in neuroscience (Rieke et al., 1997;  Borst and Theunissen, 1999). Very recently, information theory has been applied to genetic networks to study development in fruit fly embryos and to investigate properties of small stochastic biochemical networks (Tkacik et al., 2008a, 2008b; Ziv et al., 2007; Walczak et al., 2009; Tostevia and ten Wolde, 2009). Here, we adapt information theory to study a biological circuit with multiple inputs and a single output.

One of the advantages of using Information Theory to describe cellular signaling is that no detailed knowledge of the components and kinetic parameters that constitute the signaling circuit is required. Rather, the signaling circuit is modeled by its input-output relationship, often called the transfer function, which describes how the output varies as a function of the input signals. The transfer function of the {\it V. harveyi} quorum-sensing circuit was recently measured using single-cell fluorescence microscopy (see Fig. \ref{fig:fig0}b and Long et al., 2009). Below, we show that many features of the experimentally measured transfer function can be understood using information theory. We argue that our analysis of the {\it V. harveyi} quorum-sensing network provides insight into broader design principles applicable to many signal integration networks in cells.

\section{ Results and Discussion}

\subsection{Overview of the information theory formalism}

A central concept in Information Theory is mutual information. The mutual information is a symmetric measure of the correlation between inputs and outputs and measures how much one can learn on average  about the input from the output, and vice versa. Consequently, calculating the mutual information between each input signal and the output  in a multi-input  signaling system such as the {\it V. harveyi} quorum-sensing circuit allows us to quantify how much bacteria can learn on average about each input from the shared output.

There are three basic components of our information theory formalism: (1) a model of the signaling circuit, (2) a statistical model (a ``prior'')  for the likelihood of encountering a particular input signal, and (3) the various mutual informations between the inputs and the output (Fig. \ref{fig:fig1} a). We now discuss each of these components in greater detail.
\\
\\
(1) {\it Model for  the signaling circuit.} Signal integration in bacteria commonly occurs through two-component signal-transduction systems, often including a phosphorelay.  Additionally, the time-scales on which external input signals  such as autoinducers vary is much slower than the typical time scales for phosphorylation/dephosphorylation in the relay.  This separation of time scales allows us to model  the signaling circuit by its steady-state properties, with the inputs assumed to be constant in time. Furthermore, for simplicity and  to facilitate comparison with experiment, we limit our considerations to multi-input circuits with two input signals, denoted $X$ and $Y$, and a single output, denoted $Z$. The generalization to circuits with more than two inputs is straightforward.

 For an idealized multi-input channel without biochemical noise, an input $(X,Y)$ gives rise to a single output $Z=f(X,Y)$. However, signaling fidelity is generally limited by biochemical noise so that a single input can give rise to many outputs (McAdams and Arkin, 1997; Ozbudak et al., 2002; Swain et al., 2002; Elowitz et al., 2002). Noise can arise from both the stochastic nature of biochemical reactions, often called intrinsic noise, and from other cellular variability, often called extrinsic noise. 
 Whereas the former can be reduced by temporal averaging, the latter often cannot. Consequently, we characterize a signaling circuit  by a noisy transfer function, $P(Z|X,Y)$,  which gives the probability of  an output, $Z$,  as a function on the inputs, $X$ and $Y$ (cf. Fig. \ref{fig:fig1}b).\\
\\
(2) {\it Prior distribution on input signals.}   In order to quantify information transmission, it is necessary to define a prior distribution of input signals, $q(X,Y)$. This prior represents the probability that a bacterium receives an input signal $(X,Y)$.  For example, $q(X,Y)$ could be the distribution of input signals that a typical bacterium would encounter in its natural habitat (Tkacik et al., 2008a). 
\\
\\
(3){\it Mutual information.} Information transmission in a signaling circuit can be quantified by the mutual information between the input and output signals. For a circuit with two inputs, $X$ and $Y$,  and a single output, $Z$, there are  three distinct mutual informations,  $I(Z,X)$, $I(Z,Y)$, and $I(Z,(X,Y))$. They measure, respectively,  how much can be learned on average about the inputs $X$, $Y$, and $(X,Y)$, from the output $Z$.   $I(Z, (X,Y))$ measures the total information transmitted about both signals $X$ and $Y$  but is a poor measure of how much can be learned about individual inputs. By contrast,  $I(Z,X)$ and $I(Z,Y)$ measure how much can be learned about the individual inputs, but these are often not reflective of total information transmission.
 
Mutual information is a  statistical quantity that measures the {\it average} amount that can be learned  about an input, with the average taken over the input prior, $q(X,Y)$. Consequently, all three  mutual informations depend on both the transfer function of the signaling circuit, $P(Z|X,Y)$ and on the prior, $q(X,Y)$. The relevant expressions are (Shannon, 1946; MacKay, 2003)
\be
I(Z,(X,Y)) = \int dZ dX dY p(Z,X,Y) \log_2{\left(\frac{p(Z,X,Y)}{p(Z)q(X,Y)}\right)},
\ee
with the joint probability  $p(Z,X,Y)=P(Z|X,Y)q(X,Y)$ and
\\$p(Z)=\int dX dY P(Z|X,Y)q(X,Y)$, and
\be
I(Z, X)= \int dZ dX p(Z,X) \log_2{\left(\frac{p(Z,X)}{p(Z) q(X)}\right)},
\ee
with $p(Z,X)=\int dY P(Z|X,Y)q(X,Y)$ and $p(Z)$ as above, and the expression for $I(Z,Y)$ is the same as $I(Z,X)$ except with $X$ and $Y$ interchanged.  Information is measured in bits, indicated by the use of the base two logarithm in the expressions above.  A bit is the standard unit of information  and is defined as the quantity of information required to distinguish two mutually exclusive but equally probable states from each other.

\subsection{Information transmission in  the {\it V. harveyi} quorum-sensing circuit} 

Applying Information Theory to the {\it V. harveyi} quorum-sensing circuit requires explicit models for the transfer function and the prior. The input-output relationship for the {\it V. harveyi} quorum-sensing circuit was experimentally quantified in genetically engineered strains lacking the  CAI-1-CqsS pathway in order to study the integration of signals from autoinducers AI-1 and AI-2 (Long et al., 2009).  In these experiments, strains were engineered with {\it gfp} fused to the promoter of {\it qrr}4, which is one of the genes encoding the quorum-sensing small RNAs activated by phospho-LuxO. As shown in Fig. \ref{fig:fig0}, the signals from AI-1 and AI-2 are already integrated at this stage of the quorum-sensing circuit. Recent experimental and theoretical work suggest that the detection of AIs by their cognate receptors  (e.g. LuxN, LuxPQ) can be understood using a simple two-state model in which  receptors exists in two states: a low kinase activity state (``off") and a high kinase activity state (``on") (Swem et al., 2008; Keymer et al., 2006; Supporting Information). In addition to their kinase activities, the quorum-sensing receptors have a strong state-independent phosphatase activity (Long et al., 2009).  AIs act by  binding to a receptor and decreasing the probability that the receptor is in the high kinase activity, ``on'' state. Thus, specifying the external concentration of an AI  in the environment is equivalent to specifying the probability that the corresponding receptor is in its high activity state. Hence, we take the input signals, $X$ and $Y$, to be the probabilities that LuxN and LuxPQ, respectively, are in their kinase-active states. An advantage of this formulation is that input signals are bounded between $0$ and $1$ (Supporting Information).

Motivated by experiment (Long et al., 2009), we model the mean response of the {\it V. harveyi} quorum-sensing circuit using the expression 
\be
Z= f(X,Y)= \frac{k_X X + k_Y Y}{k_X X  + k_Y Y+ p} \approx  \frac{k_X}{p}  X+ \frac{k_Y}{p} Y,
\label{qsmeantf}
\ee
with $Z$ the output signal, i.e. the fraction of phospho-LuxO,  $k_X$ the total kinase rate of active LuxN, $k_Y$ the total kinase rate of active  LuxPQ, and $p$ the total phosphatase rate from both receptors. The second, approximate expression, applies because, for the quorum-sensing circuit, the total phosphatase rate is much larger than the maximal total kinase rate, $p \gg k_X + k_Y$.  The mean transfer function $f(X,Y)$ is plotted in Fig. \ref{fig:fig2} for the cases $k_X \gg k_Y$, $k_X=k_Y$, and $k_X \ll k_Y$. Experiments indicate that the actual kinase activities of the AI-1 and AI-2 pathways are nearly equal (Fig. \ref{fig:fig0}b, Long et al., 2009).  

In a standard fashion, we approximate the probabilistic transfer function, $P(Z|X,Y)$ as a Gaussian channel, in which the probability of observing an output for a given input is modeled by a Gaussian distribution around the mean output level for that input (Detwiler et al., 2005; Tkacik et al., 2008a, 2008b; Ziv et al., 2007; MacKay, 2003). Explicitly, we model the  noisy transfer function as
\be
P(Z|X,Y)= \frac{1}{\sqrt{2\pi\sigma^2(X,Y)}}\exp{\left(-\frac{[Z-f(X,Y)]^2}{2\sigma^2(X,Y)}\right) },
\label{defnoisytf}
\ee
where $f(X,Y)$, given by Eq. \ref{qsmeantf},  is the deterministic transfer function describing the average output $Z$ as a function of the inputs $X$ and $Y$, and $\sigma(X,Y)$ is the input-dependent standard deviation of the output signal for a given input (cf. Fig. \ref{fig:fig1}b). We expect this to be a good approximation because, experimentally, the noise is well approximated by a Gaussian and is much smaller than the mean signal, $\sigma(X,Y)/f(X,Y) \ll 1$(Long et al., 2009).

Unfortunately, little is known at a quantitative level about the natural environment of {\it V. harveyi}, making it difficult to accurately model the prior $q(X,Y)$.  Therefore, we take the approach of performing all our calculations for a variety of reasonable priors. In this report, we present results for three choices of prior: a flat prior where  all inputs are equally likely, a bimodal prior which is symmetric in the two inputs, and a non-symmetric bimodal prior (see Supporting Information). We have verified that our main conclusions are insensitive to the choice of prior.

\subsection{ Information about each input is limited by ``noise''  from the other input(s) } 
In a circuit that integrates multiple signals, information transmission about each individual signal is limited by two distinct phenomena, biochemical noise and interference from other signals (cf. Fig. \ref{fig:fig1}). Noise arises from both the stochastic nature of the biochemical reactions underlying the signaling circuit  and as well as other sources of cellular variability (Elowitz et al., 2002). In the presence of noise, a single input gives rise to a distribution of outputs. This type of noise limits information transmission because it introduces uncertainty about the input given the output.  A second, independent phenomenon that limits information transmission about individual inputs  in multi-input circuits is interference from other signals.  Generally, different combinations of the input signals can  give rise to the same output signal. Consequently, when a multi-input circuit is viewed as a single-input channel for a particular input, other signals introduce additional uncertainty  about that input even in the absence of noise, i.e. the other signals act as additional noise sources (cf. Fig. \ref{fig:fig1}c).

In the {\it V. harveyi} quorum-sensing circuit, experiments indicate that the noise is generally significantly smaller than the mean input signal, with the signal-to-noise ratio always greater than 2.5 ($\sigma(X,Y)/f(X,Y) \ge 2.5 $. Thus, the circuit is always in a `low-noise' regime. To assess whether noise or interference from other signals is the primary limitation on information transmission about individual signals, we have obtained formulas for the mutual informations $I(Z,X)$ and $I(Z, Y)$ in the low-noise regime using a saddle-point approximation (see Supporting Information). Recall that  $I(Z,X)$ and $I(Z,Y)$ measure the average amount of  information that can be learned about the individual inputs $X$ and $Y$ from the output $Z$, and therefore $I(Z,X)$ and $I(Z,Y)$ allow us to quantify information transmission about the individual inputs. We find that our approximate expressions for these quantities do not depend on the noise, indicating that information transmission about each input is primarily limited by interference from the other signal.

\subsection{ Total information transmission is limited by biochemical noise}

We can also discover  how much bacteria can learn on average about all the inputs from the mutual information $I(Z, (X,Y))$ between the output $Z$ and the ordered pair of inputs $(X,Y)$. In contrast to the case of individual inputs considered above, we find that even in the low-noise regime,  total information transmission is limited by noise when both signals are considered. Our approximate expressions are analogous to those obtained for a single-input, single-output biochemical network (Tkacik et al., 2008b; Supporting Information). This follows intuitively because $I(Z,(X,Y))$ is insensitive to the identity of the individual signals $X$ and $Y$ and thus the circuit effectively has a single input $(X,Y)$ and a single  output $Z$.

We calculated the total information transmission in the {\it V. harveyi} quorum-sensing circuit using data from Long et al.  2009  (see Supporting Information). We calculated the mutual information $I({\rm GFP}, (X,Y))$ between the GFP output signal and the inputs and found that the information is of order 1.5 bits for a variety of  priors. Note that by standard information theoretic inequalities,  $I({\rm GFP}, (X,Y))$  is a lower bound on $I({\rm Phospho \text{-} LuxO}, (X,Y))$ the information transmitted between the inputs and LuxO, the output of the quorum-sensing phosphorelay (MacKay, 2003). Nonetheless, we stress that  $I({\rm GFP}, (X,Y))$ is a reasonable proxy for  the true information transmission since information from the inputs is eventually transmitted to the master quorum-sensing regulator LuxR via the small RNAs (cf. Fig. \ref{fig:fig1}a).

\subsection{ {\it V. harveyi}  must tune kinase activities to simultaneously learn about multiple inputs}

Experiments indicate that in {\it V. harveyi},  signals from two of the  autoinducers, AI-1 and AI-2, are combined strictly additively in a shared phosphorelay pathway, with each autoinducer contributing very nearly equally to the total response (Long et al., 2009). In terms of the mean response  (Eq. \ref{qsmeantf}), this means that the maximal kinase activities of the AI-1/LuxN  and AI-2/LuxPQ pathways are almost identical, i.e.   $k_X \approx k_Y$. The observed transfer function appears puzzling at first -- it is symmetric in the two inputs (cf. Fig \ref{fig:fig0}b) indicating that bacteria cannot distinguish between AI-1 and AI-2 even though the two AIs encode distinct  information about local species composition. This conundrum motivated us to investigate how the kinase rates of the two pathways, $k_X$ and $k_Y$, and phosphatase rate, $p$, affect information transmission, by  calculating the mutual informations, $I(Z,X)$ and $I(Z,Y)$, for different choices of circuit parameters.

 Our results indicate that the signal processing properties of the quorum-sensing circuit vary dramatically with changes in the relative strength of the kinase activities of the two pathways. In contrast, the net phosphatase activity, $p$, affects  information transmission only modestly (data not shown). Fig. \ref{fig:fig3} shows plots of $I(Z,X)$ and $I(Z,Y)$, as functions of  the ratio of kinase activities $k_Y/k_X$ for various priors.  As discussed previously, $I(Z,X)$ and $I(Z,Y)$ are limited primarily by interference between signals, not by noise. Therefore, we used the  low-noise expressions (see Supporting Information). Our results indicate that  if $k_Y/k_X \gg1$,  $I(Z,Y)$ can be very large ($\gg 1$) but $I(Z,X)$ is very small ($\ll 1$).  One the other hand, if  $k_Y/k_X \ll 1$,  $I(Z,Y)$ is very small but $I(Z,X)$ is very large. Thus,  if the kinase activity of one pathway is much larger than the other, the cell can only  learn about the stronger pathway. Only when the kinase activities of the two pathways are roughly equal, $k_Y \approx k_X$, can the cell learn about both inputs. We conclude that {\it V. harveyi}  must tune the kinase activities of the AI-1 and AI-2 pathways to be roughly equal in order to learn about both inputs. Indeed, this is what is observed in experiments (Long et al., 2009).
 
These results can be intuitively understood as follows.  The architecture of prokaryotic phosphorelays is such that a single phosphate group is passed from the receptors detecting the inputs to the output response regulator. Thus, in the  {\it V. harveyi} quorum-sensing circuit, all information about the inputs is encoded in a single number, the number of phospho-LuxO molecules. At steady state, bacteria are limited to what is commonly referred to in the language of information theory as ``amplitude encoding''. Amplitude encoding places strong limitations on how signals can be integrated. If the kinase rate of the $X$-signaling branch is much larger than that of the $Y$-signaling branch, ($k_Y/k_X \ll 1$), then the number of phospho-LuxO  almost entirely reflects the magnitude of the input  signal $X$ and contains very little information about $Y$. On the other hand, if   ($k_Y/k_X \gg 1$), then the number of phospho-LuxO  almost entirely reflects the magnitude of the input  signal $Y$ and contains very little information about $X$. This relation can be observed graphically from the constant output contours of Fig. \ref{fig:fig2}. The contours are almost vertical when $k_Y/k_X \ll 1$ indicating $Z$ is highly correlated with $X$ but largely uncorrelated with $Y$. The opposite is true when $k_Y/k_X \gg 1$.

We conclude that in order for the number of phospho-LuxO molecules to contain information about both signaling branches, it is necessary that, on average, both signaling branches phosphorylate about equal numbers of LuxO. In terms of kinase activities, this translates into the requirement that the maximal kinase activites of the two signaling pathways be approximately equal, $k_X \approx k_Y$.
In light of these results, we speculate that the reason that  the kinase activities of the  AI-1/LuxN  and AI-2 /LuxPQ pathways are nearly identical is to allow bacteria to learn about the concentration of both autoinducers individually. However, setting $k_X \approx k_Y$  has the consequence of introducing symmetry in the input-output relation such that the bacteria cannot distinguish the input $(X,Y)$ from the input $(Y,X)$. For example, the bacteria cannot  distinguish saturating  AI-1 and no AI-2 from saturating AI-2 and no AI-1 (Long et al., 2009).  This constraint limits how much cells can learn about either input. Quantitatively, from our calculations, for $k_X \approx k_Y$,  cells can learn only about 0.75 bits about each input signal.

\subsection{ Bacteria can increase information transmission by manipulating the inputs}

In quorum sensing, bacteria both produce and detect autoinducers. This led us to consider if bacteria could increase the information they obtain about their environments by manipulating relative autoinducer production rates. As discussed,  the primary limit on information transmission when the kinase rates of the AI-1 and AI-2 signaling pathways are equal is the symmetry in the input-output relation. We hypothesized that  bacteria may distinctly manipulate the different autoinducer production rates in order to remove the ambiguity between the two input signals and thereby increase the information the AIs provide. For example, bacteria could temporally segregate signals by first producing one autoinducer and then the other. Alternatively, {\it V. harveyi} could produce AI-1 and AI-2 at the same rate. This solution would  ensure that  there was always more AI-2 than AI-1 in the environment because AI-1 is produced only by {\it V. harveyi} whereas AI-2 is produced by almost all bacteria. Within our model, this arrangement corresponds to limiting the input signaling space to $X \ge Y$. We calculated $I(Z,X)$ and $I(Z,Y)$ for the latter scenario and the results are shown in Fig. \ref{fig:fig4}. We found that when $k_X \approx k_Y$  and $X \ge Y$ bacteria could  learn $\approx 1.5$ bits  about each signal, double what they can learn when the input space is unrestricted. This result confirms that in principle, bacteria can increase information transmission by manipulating autoinducer production rates.

\subsection{Feedback on receptor number allows bacteria to focus attention on individual inputs}

The information-theory analysis presented above shows that the signaling properties of the {\it V. harveyi} quorum-sensing circuit are sensitive to changes in the kinase rates of the inputs. This raises the intriguing possibility that bacteria might implement more sophisticated signal detection strategies by varying kinase activities as a function of inputs. A simple architecture for achieving this goal would be a feedback on receptor number. Indeed, such a feedback has recently been discovered in the quorum-sensing circuit of {\it V. harveyi}, with sRNAs negatively regulating production of LuxN (Schaffer and Bassler, unpublished). We show below that such feedbacks on receptor number potentially allow bacteria to ``focus attention'' on different inputs depending on their external environments.
 
The maximal kinase activity of each autoinducer pathway in {\it V. harveyi} depends on two separate quantities:  (1) the total number of receptors, and (2) the maximal kinase activity of each individual receptor. Explicitly, the maximal kinase rates of the  $X$ (AI-1) and $Y$ (AI-2) pathways obey  $k_X=k_X^0N_X$ and $k_Y=k_Y^0N_Y$, with  $N_X$ and $N_Y$  the number of receptors in the $X$ and $Y$ pathways, respectively, and $k_X^0$ and $k_Y^0$ the maximal kinase activities of the individual receptors. Thus, in principle, bacteria can modulate the ratio of maximal kinase rates between the two pathways, $k_Y/k_X$, as a function of the output, $Z$, through feedback on receptor number ( cf. Fig. \ref{fig:fig6} and Supporting Information).  

We consider here two simple feedback architectures: (1) positive feedback on $N_Y$  and (2) negative feedback on $N_X$.  Both of these feedback architectures  allow bacteria to tune kinase rates of the two pathways  so that $k_Y/k_X \gg 1$ at large $(X,Y)$ and $k_Y/k_X \ll 1$ at small $(X,Y)$.  This can be understood graphically in Fig. {\ref{fig:fig6}a which shows contour lines of constant output, $Z$, for different values of the inputs $X$ and $Y$ in the presence of a positive feedback from $Z$ on $N_Y$. Notice that  for $X$ and $Y$  near $1$ ($i.e.$ at low cell density), the constant-$Z$ contours are more horizontal indicating that $k_Y/k_X \gg 1$ whereas for $X$ and $Y$ close to zero (i.e. high cell density), the contour lines are much more vertical indicating that $k_Y/k_X \ll 1$.  Therefore, for the feedback shown in Fig. \ref{fig:fig6}a, bacteria preferentially learn about $Y$ (AI-2) at low cell densities and about $X$ (AI-1) at high cell densities. Analagous results can be achieved using a negative feedback on $N_X$  (cf. Fig \ref{fig:fig6}b and Supporting Information).

To quantify information transmission for such feedback architectures, we calculated the mutual informations $I(Z,X)$ and $I(Z,Y)$ in the presence of feedbacks for various choices of kinetic parameters (see Supporting Information). We found that the mutual informations in the presence of either a positive feedback on the receptors for input Y or a negative feedback on the receptors for input  X are comparable to those in the absence of feedback. This finding indicates that bacteria can preferentially detect AI-2 ($Y$) at low cell densities and AI-1 ($X$) at high cell densities without sacrificing how much they learn on average about both inputs. For example, for both the feedback transfer functions shown in Fig. \ref{fig:fig6}, $I(Z,X)$ and $I(Z,Y)$ are both $\approx$ 1.5 bits for the case when $X \ge Y$, comparable to their values in the absence of feedback (cf. Fig, \ref{fig:fig3}).

\subsection{ Discussion}

Cells constantly sense their environments and adjust their behavior accordingly.
Specifically, cells often integrate temporally coincident information from multiple environmental inputs to modulate their gene expression states. However, the mechanisms and logic by which cells integrate
multiple signals remain by and large poorly understood. We developed a new, mathematical framework for analyzing information processing in cells based on Information Theory and used it to study the integration of multiple autoinducer signals by the model quorum-sensing bacterium {\it Vibrio harveyi}. Our studies revealed that there are two distinct mechanisms that limit information transmission when bacteria integrate multiple signals, biochemical noise and interference between different signals. Whereas the former limits the total information that bacteria can learn about all the inputs, signal interference is the primary impediment to learning about individual input signals. Furthermore, we showed that because of signal interference, {\it V. harveyi} cells must precisely tune the kinase activity of each input branch of the quorum-sensing pathway in order to simultaneously learn about individual autoinducer inputs. These theoretically motivated conclusions are consistent with recent quantitative experiments on {\it V. harveyi} showing that the maximal kinase activities of the AI-1 (LuxN) and AI-2 (LuxPQ) pathways are nearly equal (Long et al., 2009). Our information-theory analysis also indicates that bacteria can increase how much they learn about individual inputs by manipulating the different autoinducer production rates. Finally, we have shown that bacteria can learn preferentially about a particular signal in a particular environment, even with a single-output pathway, by using simple feedback loops to control receptor numbers.

Our theory not only explains the puzzling experimental observation of nearly equal kinase activities of the LuxN and LuxPQ pathways (Long et al., 2009), but also makes several testable predictions about the {\it V. harveyi} quorum-sensing circuit. First, we predict that the maximal kinase activity of the  CAI-1/CqsS branch, when measured, will prove to be similar to that of the AI-1/LuxN and AI-2/LuxPQ pathways (see Fig. \ref{fig:fig1}). This  prediction follows directly from our information-theory analysis which indicates that the three signaling branches must phosphorylate about equal numbers of LuxO in order for cells to simultaneously learn about all three input signals. Second, the theoretical work presented here suggests that {\it V. harveyi} may manipulate both autoinducer production and receptor numbers in order to reduce interference between signals and thereby increase information transmission. Preliminary evidence suggest that this is the case.

An as yet unanswered question is why {\it V. harveyi} and related species employ multiple autoinducers (AIs) but then funnel all the information from these autoinducers into a single-output pathway. We speculate that different concentrations of multiple autoinducers may represent different stages of community development such as the stages of growth in a biofilm. Unlike eukaryotic development, $e.g.$ embryogenesis, where  the rate of development is fixed and driven by a clock (Nieuwkoop and Faber, 1994) and the input signal is often stereotyped (Gregor et al., 2007a, 2007b), the rate of development of a bacterial community depends on unpredictable environmental conditions such as nutrient availability and population composition and density. To compensate for such variability, quorum sensing could allow bacteria to monitor stages of community development and act accordingly. The architecture of the {\it V. harveyi} quorum-sensing circuit, with multiple inputs and  a single output, is consistent with the idea that  {\it V. harveyi} uses quorum sensing to implement a single, multi-stage developmental program. Indeed, Long et al. (2009) showed that {\it V. harveyi} can ``count''  the number of autoinducer signals present. Thus, if AIs accumulate in a defined sequential order, the number of autoinducers present at saturating concentration could signal different stages of development. For example,  models of biofilm growth suggest that the universal autoinducer AI-2 may be more informative at early stages of biofilm growth where communities are expected to be mixed whereas the species specific autoinducer  AI-1 may be more informative at later stages when mostly progeny are nearby (Nadell et al, 2007). 

Our detailed analysis of the {\it V. harveyi} quorum-sensing network has implications for other prokaryotic signal-integration networks. Signal integration is a  common feature of many organisms, and bacteria have developed sophisticated molecular mechanisms for integrating signals from a broad range of inputs using two-component systems and phosphorelays (Kato et al., 2007;  Mitraphanov and Groisman, 2008; Bassler and Losick, 2006; Perego, 1998). For example, the sporulation and competence circuits of the soil-dwelling bacterium {\it Bacillus subtilis} integrate signals from the environment, cell-cycle, and metabolism using a network design based on competition between various protein kinases and phosphatases (Perego, 1998; Veening et al., 2008).  Our information-theory analysis suggests that the need to minimize interference between signals likely places strong constraints on the design of such signal-integration networks. In particular, our work indicates that the information transmission properties  are likely to be extremely sensitive to changes in kinase and phosphatase rates, and that bacteria may have evolved strategies for minimizing interference. One possible strategy for learning about individual input signals is to temporally coordinate signals (Mitraphanov and Groisman, 2008).  For example, recent experiments on the {\it B. subtilis} sporulation and competence networks indicate that bacteria likely temporally separate input signals (Smits et al., 2007; Veening et al., 2008). 

Bacteria may employ a range of mechanisms to minimize signal interference by actively controlling both signal production and detection. A simple way to achieve such control is via feedbacks on synthases/receptors. For example, feedbacks on AI production are a common feature of many quorum-sensing systems (Waters and Bassler, 2005), suggesting bacteria actively manipulate the temporal profile of AI production. In {\it V. harveyi}, such a feedback has recently been found to act on the AI-1 synthase LuxM (Schaffer and Bassler, unpublished). Moreover, the gene encoding LuxM  is located in an operon with the gene encoding the AI-1 receptor LuxN indicating the feedback also acts on receptor numbers, potentially allowing {\it V. harveyi} to focus on different autoinducers at different stages of development. Recent experiments also indicate that {\it E. coli} cells manipulate chemoreceptor numbers using feedbacks. When starving, {\it E. coli} change the ratio of Tar to Tsr receptors, resulting in a change of behavior from heat seeking to cold seeking (Salman and Libchaber, 2007).

The use of signaling pathways with multiple inputs and a single output necessarily entails a loss of information about  input signals. This raises the natural question of why such pathways are utilized by bacteria. For example, one can imagine alternative architectures where each input is detected by a dedicated  signaling pathway and information about multiple inputs is integrated at the promoters of regulated genes via combinatorial gene regulation. We have argued that for {\it V. harveyi} such a multi-input, single-output architecture facilitates the implementation of a linear, multi-stage,  developmental program.  The architecture of signal integration networks may also reflect evolutionary constraints. For example, such networks may have evolved from a single pathway by gene duplication. Additionally, when the output of a signal-integration network is a master transcription factor regulating the expression of many genes (e.g. LuxR in {\it V. harveyi}), the use of  a single-output pathway may be more efficient with regard to use of space on the genome than a competing architectures consisting of individual signaling pathways, one for each input, culminating in combinatorial gene regulation. Recent experiments indicate that the sporulation network in {\it B. subtilis} may play a similar role in regulating  biofilm formation (Vlamakis et al., 2008). In light of the accumulating evidence that  bacterial populations behave similarly to multicellular organisms (Shapiro, 1998), we suspect that the use signal-integration networks to coordinate development programs may be widespread in prokaryotes. 

The work presented here focuses on signal integration in bacteria. The architecture of prokaryotic phosphorelays, where a single phosphate group is transferred sequentially  to downstream components, constrains bacteria to encode information using amplitude encoding, i.e. all information about input signals is contained in the number of active response regulator molecules. The use of amplitude encoding places strong constraints on network architecture and limits the amount of information that bacteria can transmit. This contrasts with neural networks, where spike timing allows neurons to encode information using more sophisticated schemes (Reike et al.,1997). Signaling in bacteria also differs from signaling in eukaryotes, which often utilizes multiple phosphorylation sites and kinase cascades that permit temporal encoding schemes such as dose-duration encoding (Behar et al.,2008; Detwiler et al., 2000).  It may prove fruitful to generalize our information theoretic formalism to these more complicated intracellular circuits. 

 Finally, our results suggest that Information Theory may prove to be a powerful general tool for analyzing biological signaling networks. Information theory provides a natural language for formulating questions about information processing and signaling integration. An additional advantage of an information-theoretic analysis is that no detailed knowledge of the signaling circuit is required. All quantities are calculated using the input-output relationship of a signaling circuit, often an experimentally accessible quantity, even for large signaling networks. For these reasons, we expect the application of Information Theory to yield novel biological insights into cellular signaling in the future. 
 
\section{Acknowledgements} We would like to thank Thierry Mora, Anirvan Senputa, the Wingreen and  Bassler labs, and the Princeton biophysics theory group for useful discussions. P.M. was supported by U.S. National Institute of Health (NIH) grant K25 GM086909-01. S.G and T.L. were partially supported by the Burroughs Wellcome Fund Graduate Training Program. S.G. and  N.W. were partially supported by the Defense Advanced Research Projects Agency (DARPA) under grant HR0011-05-1-0057. This work was also partially supported by National Science Foundation (NSF) Grant Phys-0650617.

\newpage

\section{Figure Captions}
\begin{figure}[h]
\caption { Information theoretic approach to signal integration in the {\it Vibrio harveyi} quorum-sensing circuit. (a) {\it V. harveyi} produces three distinct quorum-sensing signaling molecules, called (AIs), which are all detected by a single phosphorelay circuit that controls expression of downstream target genes. Each signaling molecule, AI-1 (red hexagons), AI-2 (blue ovals), and CAI-1 (gray squares), is detected by a cognate receptor. The receptors phosphorylate a shared phosphorelay protein, LuxU, which in turn phosphorylates LuxO. In the absence of AIs, LuxO is phosphorylated and activates  expression of  genes encoding five small regulatory RNAs (sRNAs ) which work in conjuction with Hfq to destabilize the mRNA of LuxR, the master  regulator of quorum-sensing genes. In the presence of the AIs, LuxO is not phosphorylated, the sRNAs are not produced, and LuxR is expressed. (Inset) The receptors can exist in two states: a kinase ``on" state and kinase ``off" state with ligand binding favoring the ``off" state.  (b)  Dose-response surface of  {\it V. harveyi}  to various combinations of AI-1 and AI-2  as measured in Long et al. (2009). Each vertex of the grid is the averaged normalized GFP fluorescence intensity obtained from a population of 100 cells exposed to the specified AI-1 and AI-2 concentrations  using a  {\it qrr}4{\it -gfp} transcriptional reporter fusion that is activated by phosphorylated LuxO.}
\label{fig:fig0}
\end{figure}

\begin{figure}[h]
\caption{ (a) The mutual informations $I(Z,X)$, $I(Z,Y)$, and $I(Z,(X,Y))$  measure how much one can learn about the inputs, e.g. autoinducer levels,  $X$, $Y$, and $(X,Y)$, respectively from the output $Z$, e.g. LuxR level. Mutual information is a function of  the prior, $q(X,Y)$, i.e. the a priori  probability of a given input $(X,Y)$, and of the probabilistic transfer function $P(Z|X,Y)$ of the signaling circuit. (b) (Top) For an idealized multi-input channel without noise, an input $(X,Y)$ gives rise to a single output $Z$. (Middle) In the presence of noise, a single input can give rise to many outputs with a distribution described by the noisy-transfer function, $P(Z|X,Y)$. (Bottom) When viewed as single-input channel with input $X$ and output $Z$, the second signal, $Y$, effectively acts as an additional source of noise. }
\label{fig:fig1}
\end{figure}

\begin{figure}[h]
\caption{Constant-output $Z$ contours for different relative kinase strengths of the two signaling branches carrying the inputs $X$ and $Y$: (a)  $k_Y/k_X=1/8$, (b)  $k_Y/k_X=1$, (c)  $k_Y/k_X=8$.}
\label{fig:fig2}
\end{figure}

\begin{figure}[h]
\caption{Mutual information as a function of the ratio of kinase strengths, $k_Y/k_X$. (a-c) Mutual information $I(Z,X)$ between $Z$ and $X$ (red curves), and mutual information $I(Z,Y)$ between $Z$ and $Y$, $I(Z,Y)$ (dashed blue curves) as functions of $k_Y/k_X$, for  (a) a flat prior, (b) a symmetric bimodal prior, and (c) a non-symmetric bimodal prior. (Insets) Graphical representations of the corresponding priors with brighter colors representing higher probability. }
\label{fig:fig3}
\end{figure}

\begin{figure}[h]
\caption{(a-c) Mutual information as in Fig. \ref{fig:fig3} with the restriction that the inputs obey $X \ge Y$.}
\label{fig:fig4}
\end{figure}

\begin{figure}[h]
\caption{ Graphical representation of input-output relations in the presence of a feedback on receptor number. (a) Equally spaced, constant-output $Z$ contours for a signaling circuit with positive feedback on receptor number (see inset). Parameters are $K=6$, $C=1/8$, $\delta C =8$ (see Supporting Information for definition of parameters).  (b) Equally spaced, onstant-output $Z$ contours for a signaling circuit with negative feedback on receptor number (see inset).  Parameters are $K=0.7$ and $D=d$ (see Supporting Information for definition of parameters). }
\label{fig:fig6}
\end{figure}

\newpage

\includegraphics[width=0.9\textwidth]{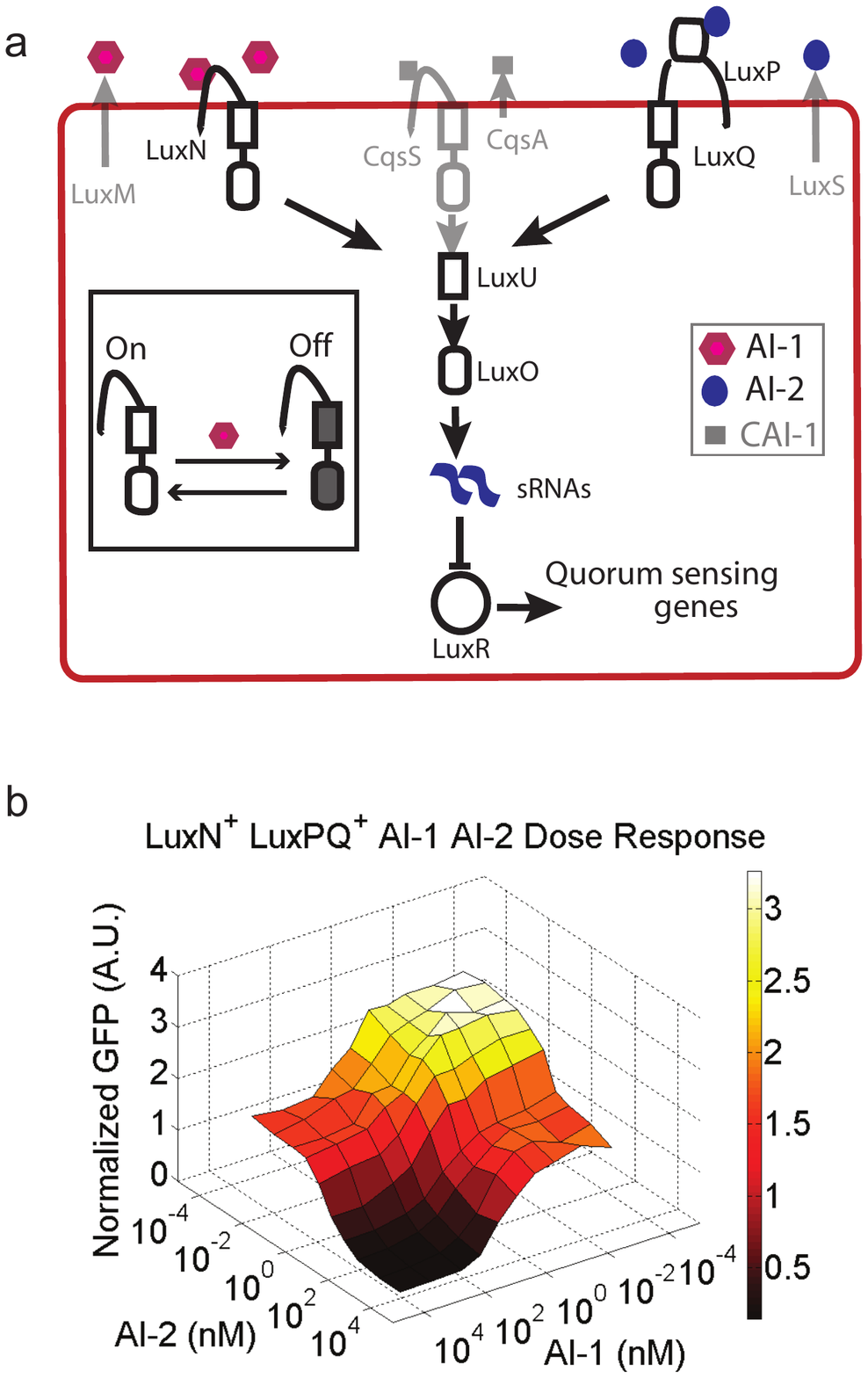}

\newpage

\includegraphics[width=0.9\textwidth]{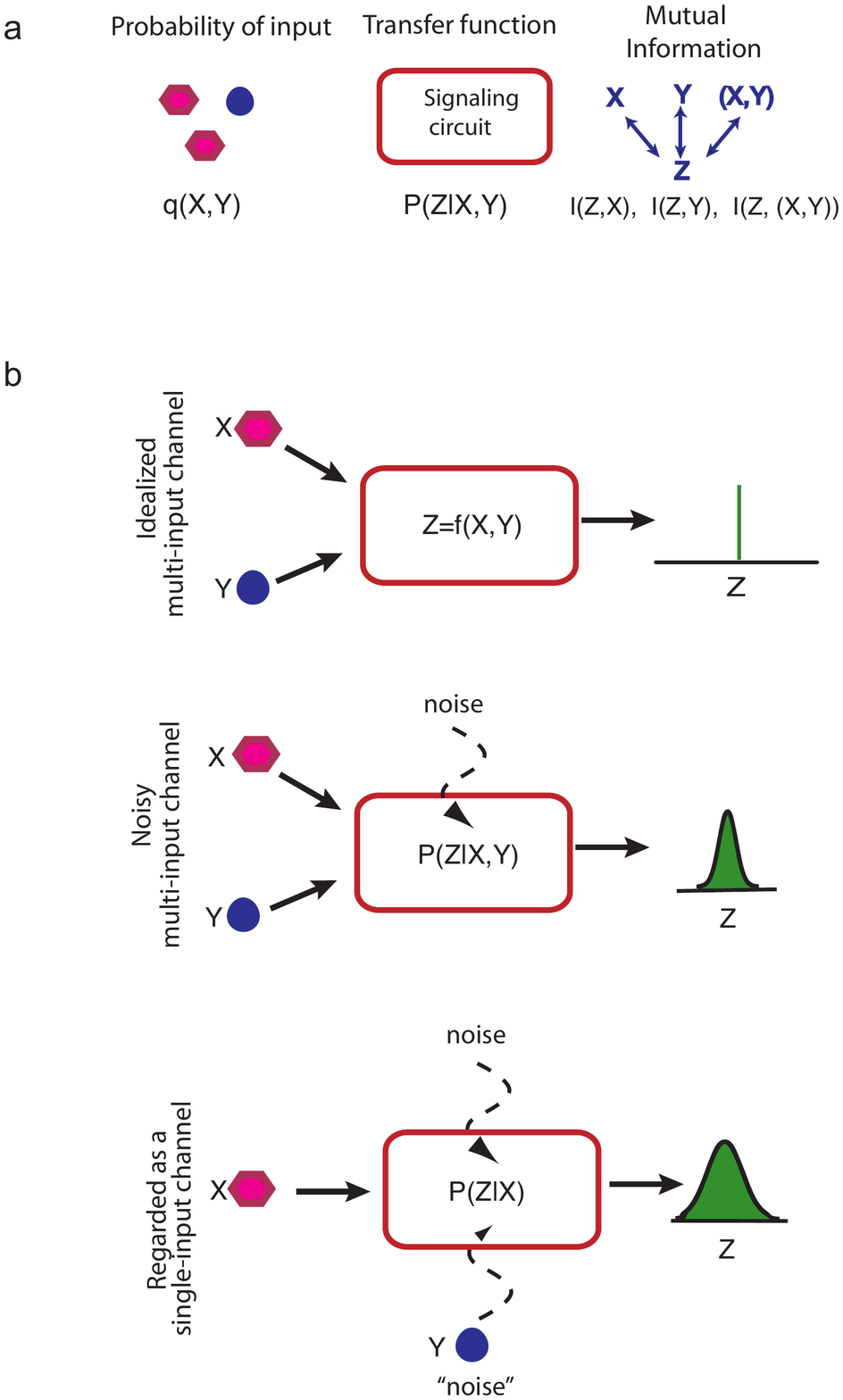}

\newpage

\includegraphics[width=0.9\textwidth]{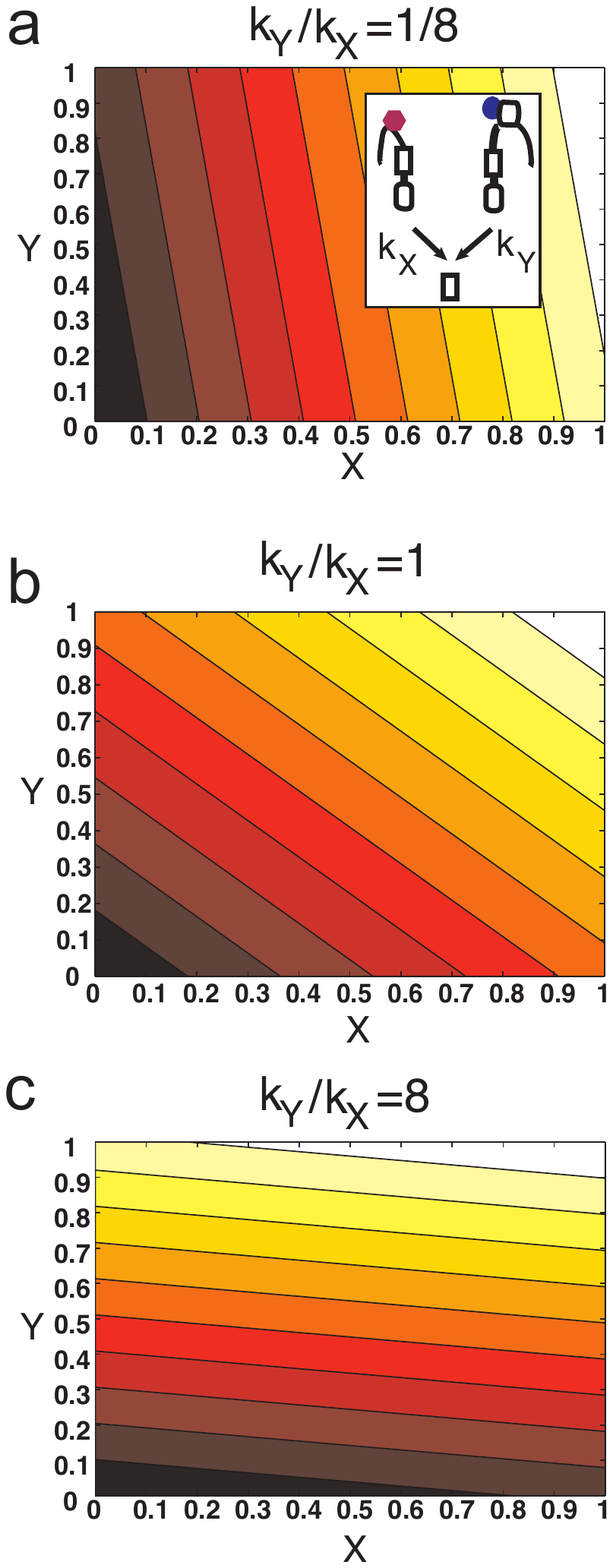}
\newpage

\includegraphics[width=0.9\textwidth]{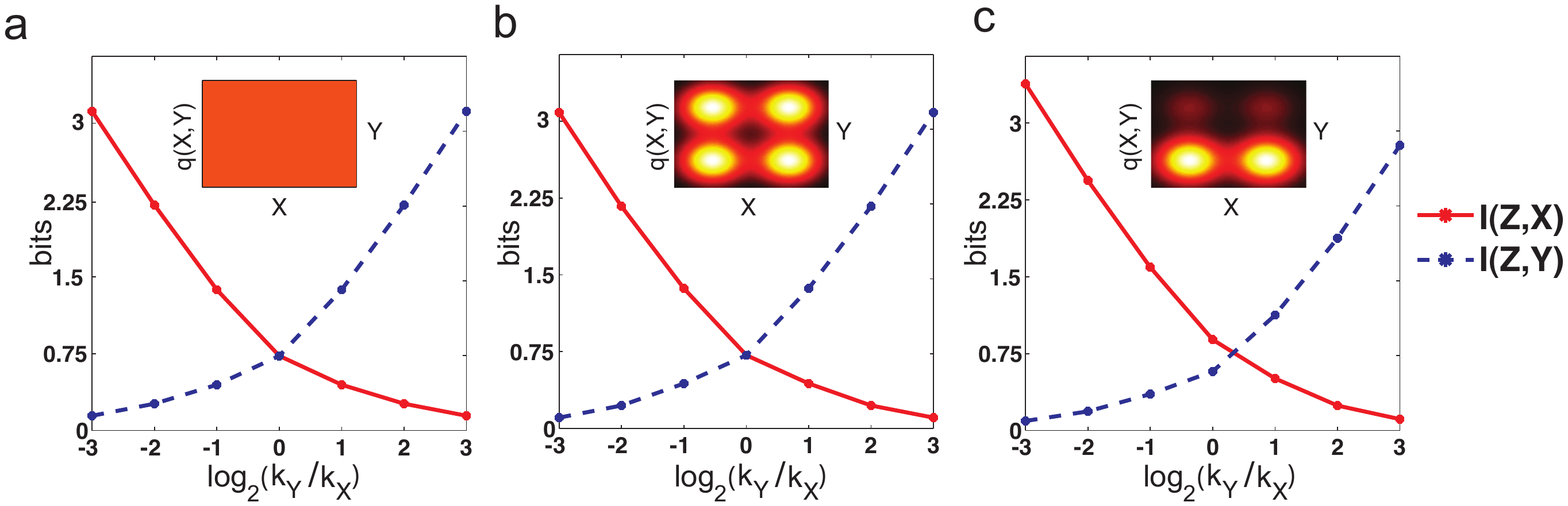}

\newpage

\includegraphics[width=0.9\textwidth]{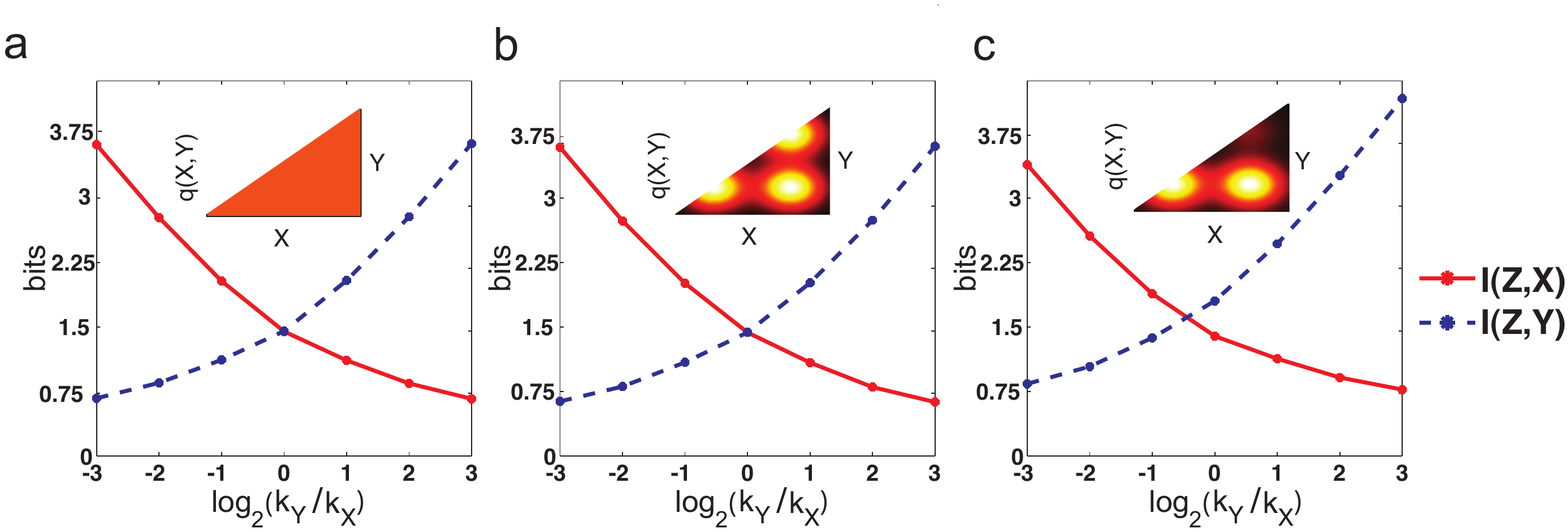}

\newpage

\includegraphics[width=0.9\textwidth]{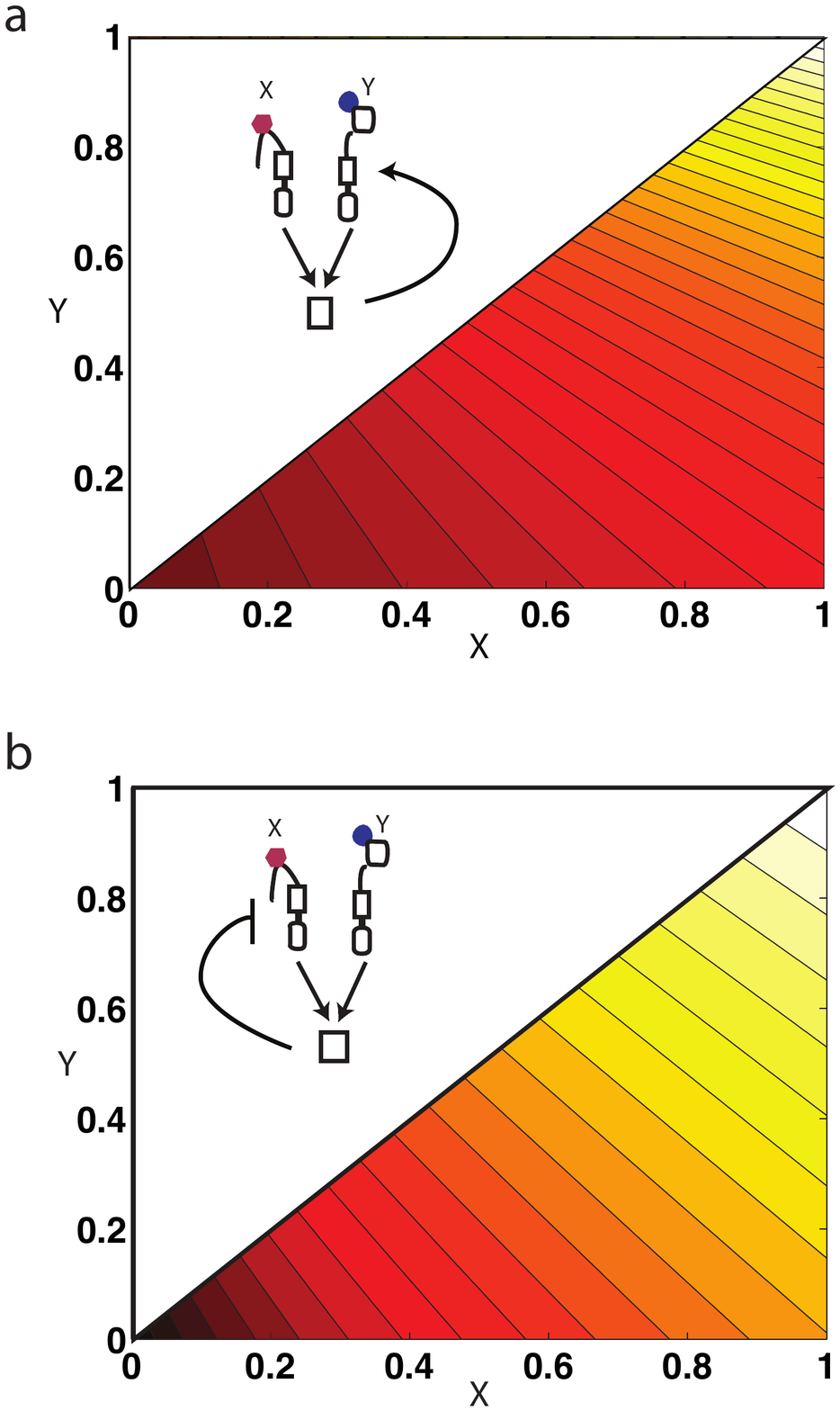}

\newpage
\appendix

\section{Two-state model for receptors}

We model receptors using a simple two-state model in which  receptors exist in two states: a low kinase activity state we call ``off" and a high kinase activity state, we call ``on" (Swem et al., 2008;  Keymer et al., 2006).  Ligands, in our case autoinducers, act by binding to the receptor protein and changing the free energies and therefore the thermal occupancies of the two activity states. There are a total of four free-energy states with corresponding free energies: $(i)$ on without ligand-bound $E^{\rm on}$, $(ii)$ on with ligand-bound $E^{\rm on} - \log{([{\rm L}] / K^{\rm on})}$, $(iii)$ off without ligand-bound $E^{\rm off}$, and (iv) off with ligand bound $ E^{\rm off} - \log{([{\rm L}] / K^{\rm off})}$. In the absence of ligands, the receptors favor the on state but ligand causes switching to the off state. This implies that $K^{\rm on} \gg K^{\rm off}$ and that $E^{\rm on} < E^{\rm off}$.  At equilibrium, the probability that a receptor is in the on state is a function of the difference in free energies between the ``on" state and the ``off" state:
\be
 f= \epsilon+  \log{\left( \frac{1+ [{\rm L}]/K^{\rm off}}{1+ [{\rm L}]/K^{\rm on}}\right)}
\ee 
with $\epsilon=E^{\rm on} - E^{\rm off}$ where all energies are expressed in units of the thermal energy $k_BT$. In particular, one has 
\be
p_{\rm on} = \frac{1}{1+ e^{f}}.
\ee
For $K^{\rm on}$ much larger than the typical ligand concentration and $\epsilon$ large and negative as in the quorum-sensing network (Swem et al., 2008), the probability that a receptor is on becomes
\bea
p_{\rm on} &=& \frac{1}{1+ e^\epsilon \left(\frac{1+ [{\rm L}]/K^{\rm off}}{1+ [{\rm L}]/K^{\rm on}}\right)} \nonumber \\
		 &\approx& \frac{1}{1+ \frac{[{\rm L}]}{K^{\rm off}e^{-\epsilon}}}.
\eea
Defining a half-maximal inhibition constant $K_I=K^{\rm off}e^{-\epsilon}$, one has the simple non-cooperative Hill function,
\be
p_{\rm on} \approx \frac{1}{1+ \frac{[L]}{K_I}}
\label{recprobon}
\ee

We denote the probabilities that  LuxN and LuxPQ  are in their on states by $X$ and $Y$, respectively. 
and we denote the kinase activities in the on states of the two receptors by $k_X$ and $k_Y$, respectively. Furthermore, for notational simplicity and consistent with experiment, we assume that the kinase activity in the off state for both receptors is negligible. We also assume, based on experimental evidence, that the receptors have state-independent phosphatase activities, which we denote $p_X$ and $p_Y$.  Since both receptors phosphorylate the response regulator LuxU, at steady state the fraction of LuxU that is phosphorylated takes the simple form
\be
\frac{\rm [LuxU\text{-}P]}{\rm [LuxU]} = \frac{k_X X+ k_Y Y}{k_x X+ k_yY +p}
\ee
where $p=p_X+p_Y$. A very similar expression can be derived for  the fraction of phosphorylated LuxO, which we denote $Z$ in the main text. We can compare these expressions to experiments in (Long et al., 2009)  by noting that from (\ref{recprobon})
\be
X  \approx \frac{1}{1+ \frac{[{\rm AI-1}]}{K_I^{\rm AI-1}}}
\ee
and
\be
Y \approx \frac{1}{1+ \frac{[{\rm AI-2}]}{K_I^{\rm AI-2}}}.
\ee

\section{Formulas for priors used in main text}

The lack of knowledge about the ecology of {\it V. harveyi} makes it difficult to quantitatively define a prior for input signals. Therefore, as discussed in the main text, we performed our calculations for several different choices of priors and verified that our conclusion are essentially independent of our choice of prior. We present results for three different choices of priors: a flat prior, a symmetric bimodal prior, and a non-symmetric bimodal prior. As discussed in the main text, we take as our inputs $X$ and $Y$ the probabilities that LuxN and LuxPQ, respectively, are in their kinase-active states. The advantage of this formulation is that input signals are bounded to be between $0$ and $1$. Explicitly, the priors we used are given by the expressions:\\
{\it Flat prior:}\\
\be
q(X,Y)=1/N
\ee
with $N$ a normalizing constant equal to 1.
\\
{\it Symmetric bimodal prior:} \\
\be
q(X,Y)= \frac{1}{N_s} \left( e^{-\frac{(X-\bar{X}_1)^2}{\sigma^2}} + e^{-\frac{(X-\bar{X}_2)^2}{\sigma^2}} \right) \left( e^{-\frac{(Y-\bar{Y}_1)^2}{\sigma^2}} + e^{-\frac{(Y-\bar{Y}_2)^2}{\sigma^2}} \right)
\label{sbmprior}
\ee
with $\bar{X}_1=\bar{Y}_1=0.25$, $\bar{X}_2=\bar{Y}_2=0.75$, $\sigma=0.2$, and $N_s$ a normalizing constant to ensure the integral of $q(X,Y)$ is one.
\\
{\it Nonsymmetric bimodal prior:}
\be
q(X,Y)= \frac{1}{N_{ns}} \left( e^{-\frac{(X-\bar{X}_1)^2}{\sigma^2}} + e^{-\frac{(X-\bar{X}_2)^2}{\sigma^2}} \right) \left( Ae^{-\frac{(Y-\bar{Y}_1)^2}{\sigma^2}} +e^{-\frac{(Y-\bar{Y}_2)^2}{\sigma^2}} \right),
\ee
with all parameters as above in Eq. (\ref{sbmprior})  plus the asymmetry parameter $A=5$, and  the normalizing constant $N_{ns}$ chosen so that the integral over the distribution is $1$.

In the last section of the main text, we restrict our input space so that $X \ge Y$. For this calculation, we use priors on the lower-half triangle of the form $q_{\rm half}(X,Y)= q(X,Y)\theta(X-Y)/N_h$ where $q(X,Y)$ is as above, $\theta(X)$ is the Heaviside function, and $N_h$ is a normalizing constant that ensures the integral over $q_{\rm half}(X,Y)$  is $1$.

\section{Mutual information via saddle-point}

\subsection{Justification for saddle-point approximation}

In the low-noise regime, we can derive approximate expressions for mutual information using a saddle-point approximation. As in all saddle point approximations, we exploit a large parameter. In our case, the large parameter is the signal-to-noise ratio. We interpret the mean value $f(X,Y)$ as the signal and $\sigma(X,Y)$ as the noise around the signal. When the noise is small, or equivalently the signal-to-noise ratio is high, we know that  $\frac{f(X,Y)}{\sigma(X,Y} \gg 1$. Thus we can write 
 $\frac{f(X,Y)}{\sigma(X,Y} = \lambda S(X,Y)$, where $\lambda \gg 1$ is a constant of order the signal to noise ratio and $S(X,Y)$ is a function of order 1. In the calculation below, $\lambda$ serves as the implicit large parameter. This implies that the saddle-point approximation is valid as long as signal-to-noise is much larger than $1$.

\subsection{Approximate probability distributions}

Often, the mean transfer functions of biological signaling systems are monotonic in the inputs.  This is true for the {\it V. harveyi} quorum-sensing circuit. In this case, it is useful to reparameterize the space of input signals in order to perform calculations. In particular, we will utilize two different coordinate systems given by the  coordinate transforms: $(X,Y) \rightarrow (r=f(X,Y), \theta=Y)$ and $(X,Y) \rightarrow (f=f(X,Y), \theta=X)$. For these coordinate transforms, by definition, we have, respectively, 
\be
q(f, \theta)= |\frac{\partial f}{\partial Y}|^{-1}q(X,Y)
\ee
and
\be
q(f, \theta)= |\frac{\partial f}{\partial X}|^{-1}q(X,Y)
\ee
where, for simplicity, we denote all distributions by the same symbol $q$ whether they are a function of $X$ and $Y$ or $f$ and  $\theta$.
By definition one has, 
\bea
p(Z)  &=& \int df d\theta \ p(Z|(f,\theta))q(f, \theta) \nonumber  \nonumber \\
          &=&\int df d\theta \ \frac{1}{\sqrt{2\pi \sigma^2(f,\theta)} }e^{-\frac{(Z-f)^2}{2\sigma^2(f,\theta)}} q(f,\theta)\nonumber \\
          &\approx & \int d\theta \ q(Z,\theta)
\eea
where, to obtain the last line, we performed the saddle-point approximation.
Furthermore,we  define the probability distributions
\be
q(\theta)= \int df  \ q(f,\theta)
\ee
and
\be
p(Z, f, \theta)= p(Z|f, \theta)q(f, \theta)= \frac{1}{\sqrt{2\pi \sigma^2(f,\theta)} }e^{-\frac{(Z-f)^2}{2\sigma^2(f,\theta)}} q(f,\theta).
\ee
A final distribution of interest to us is $p(Z, \theta)$ given by
\bea
p(Z, \theta) &=&  \int df  \ p(Z, f, \theta) \nonumber \\
&=& \int df  \ \frac{1}{\sqrt{2\pi \sigma^2(f,\theta)} }e^{-\frac{(Z-f)^2}{2\sigma^2(f,\theta)}} q(f,\theta) \nonumber \\
&\approx&  q(f, \theta).
\eea
Again, to obtain the last line, we have utilized the saddle-point approximation.

\subsection{Calculation of relevant Shannon entropies}

To calculate the mutual informations, we need several entropies: 
\bea
H(Z) &=& -\int dZ \ p(Z)\log_2{p(Z)} = -\int dZ d\theta \ q(Z, \theta) \log_2{\left[\int d\theta^\prime q(Z, \theta^\prime)\right]} \nonumber  \\
H(\theta) &=& -\int d\theta \ q(\theta) \log_2{q(\theta)} = -\int d\theta df \ q(f, \theta) \log_2{\left[ \int df^\prime q(f^\prime, \theta) \right]} \nonumber \\ 
H(Z, \theta) &=&  -\int d\theta dZ \, q(Z, \theta) \log_2{ q(Z, \theta) }.
\label{priorentropy}
\eea
A final entropy of interest to us is the entropy $H(z, r, \theta)$. Once again, we use the saddle-point approximation to obtain this entropy. Namely, one has
\bea
H(Z, f, \theta) &=& -\int dZ df d\theta \,  \frac{1}{\sqrt{2\pi \sigma^2(f,\theta)} }e^{-\frac{(Z-f)^2}{2\sigma^2(f,\theta)}} q(f,\theta) \log_2{\left[  \frac{1}{\sqrt{2\pi \sigma^2(f,\theta)} }e^{-\frac{(Z-f)^2}{2\sigma^2(f,\theta)}} q(f,\theta) \right]} \nonumber \\
&\approx&  -\int df d\theta q(f, \theta)\left[ \log_2{q(f, \theta)} +\log_2{ \frac{1}{\sqrt{2\pi}e \sigma(f,\theta) } }\right] \nonumber \\
&=& H(f, \theta) - \langle \log_2{\frac{1}{\sqrt{2\pi}e \sigma(f,\theta) }} \rangle_{q(f, \theta)},
\label{entropyall}
\eea
where, in the second line, we utilized the saddle-point approximation, and the second term in the last line is the expectation value of the logarithm of the standard deviation of the noise.

\subsection{Expressions for the individual inputs}

We calculated the information $I(Z, \theta)$. By definition,
\be
I(Z, \theta) = H(Z) + H(\theta) - H(Z, \theta).
\ee
We can use the formulas for these entropies from above to obtain the expression
\be
I(Z, \theta)= \int dZ d\theta \, q(Z, \theta) \log_2{ \frac{q(Z, \theta)}{\left[\int d\theta^\prime q(Z, \theta^\prime)\right] \times \left[\int dZ^\prime q(Z^\prime, \theta)\right]}}.
\label{priorthetainfo}
\ee
From this formula, we can calculate the information theoretic quantities of interest to us,  $I(Z, X)$ and $I(Z, Y)$ by utilizing the two different coordinate transforms discussed above: $(X, Y) \rightarrow (r=f(X,Y), \theta=X)$ and $(X, Y) \rightarrow (r=f(X,Y), \theta=Y)$. From these transforms, we know that $I(Z, \theta)$ is simply $I(Z,X)$ or $I(Z, Y)$ respectively. Note that these expressions are independent of $\sigma(f,\theta)$ and thus do not depend on the noise in the system.

\subsection{Expression for the total information}

We now calculate the total mutual information $I(Z, (f, \theta))$ between the output $Z$ and the individual inputs $X$ and $Y$. This mutual information can be expressed in terms of the entropies as
\be
I(Z, (f, \theta))= H(Z) + H(f, \theta) - H(Z, f, \theta).
\ee
Use of (\ref{entropyall}) yields the following simple expression,
\be
I(Z, (r, \theta))=  \langle \log_2{\frac{1}{\sqrt{2\pi}e \sigma(r,\theta) } }\rangle_{q(r, \theta)}  + H(Z),
\label{priortotinfo}
\ee
where $H(Z)$ is given in (\ref{priorentropy}). Since information is invariant under coordinate transforms one has $I(Z,(X,Y))= I(Z, (r, \theta))$. This expression is analogous to that found for the case of circuit with one input and one output (Tkacik et al., 2008). This follows intuitively because $I(Z,(X,Y))$ is insensitive to the identity of the individual signals $X$ and $Y$ and thus the circuit effectively has a single input $(X,Y)$ and a single  output $Z$.

\section{ Calculating total information transmission from experimental data}

We calculated total information transmission in the {\it Vibrio harveyi} quorum-sensing circuit using data from Long et al. (2009) for a variety of priors. In particular,  we calculated the mutual information between the GFP output signal and the inputs, $I({\rm GFP}, (X,Y))$ and found that it is of order $1.5$ bits for most reasonable priors. 

Here we outline our basic procedure. In Long et al. (2009), single-cell measurements were performed for a ten by ten grid of values in the $X-Y$ plane. We calculated the mean GFP level, $f(X,Y)$ as well as the variance of the GFP, $\sigma(X,Y)$,  from the data for each of these points. We susbsequently used these data to infer $f(X,Y)$ and $\sigma(X,Y)$ for all values of $X$ and $Y$ between $0$ and $1$ using quadratic interpolation.  Next, we calculated the noisy transfer function $P({\rm GFP}|X,Y)$ using Eq. 1 in the main text:  
\be
P(Z|X,Y)= \frac{1}{\sqrt{2\pi\sigma^2(X,Y)}}\exp{\left(-\frac{(Z-f(X,Y))^2}{2\sigma^2(X,Y)}\right) }
\label{totinfotranfun}
\ee
From the transfer function (\ref{totinfotranfun}), we constructed the distributions $p(Z,X,Y)$ and $p(Z)$ for various priors using the formulas
\be
p(Z,X,Y)= p(Z|X,Y)q(X,Y).
\ee
and
\be
p(Z)=\int dX dY p(Z,X,Y).
\ee
We then used these formulas and the definition of total information to obtain
\be
I(Z,(X,Y)) = \int dZ dX dY p(Z,X,Y) \log_2{\left(\frac{p(Z,X,Y)}{p(Z)q(X,Y)}\right)}.
\ee
We found that for nearly all priors $I(Z,(X,Y))$ was between 1.2 and 1.7 bits.

\section{Feedback on Receptor number}

Bacteria can manipulate receptor kinase rates using feedbacks on receptor numbers. In general, the maximal kinase activity of a pathway depends on two separate quantities: (1) the total number of receptors, and (2) the kinase activity of a single receptor. Explicitly, the maximal kinase rates of the  $X$ (AI-1) and $Y$ (AI-2) pathways of {\it V. harveyi} obey  $k_X=k_X^0N_X$ and $k_Y=k_Y^0N_Y$, with  $N_X$ and $N_Y$  the number of receptors in the $X$ and $Y$ pathways, respectively, and $k_X^0$ and $k_Y^0$ the maximal kinase activities of single receptors. Consequently, bacteria can modulate the maximal kinase activity of a pathway by changing the number of receptors using a feedback.

\subsection{Positive Feedback on Receptors}
We first consider a positive feedback on the receptors in the $Y$ pathway. In this case, the transfer function, $Z=f_{\rm fb}(X,Y)$, describing the output signal (the fraction of phosphorylated output regulators),  as a function of the inputs $X$ and $Y$ (the probability that the corresponding receptors are in their on states) is obtained by solving for the steady state of the differential equations
 \bea
 \frac{dZ}{dt} &=& (k_X X + k_Y Y) (1-Z) - p Z \nonumber \\
&=& (k_X X + k_Y^0 N_Y Y)(1-Z)-pZ, \nonumber \\
\tau \frac{dN_Y}{dt}&=&  N_{Y0} + \frac{\delta N_{Y} Z}{K+Z} -N_Y,
 \label{posfbeq1}
  \eea
where $N_{Y0}$ is the number of receptors in the in absence of feedback, $\delta N_{Y}$ measures the strength of the feedback, $K$ is value of $Z$ for which the half-maximal for the feedback. When $\delta N_y \gg N_{Y0}$, this implies that at high $Z$, $N_Y \gg N_{Y0}$. For simplicity, we have assumed a Hill coefficient  of $1$ for the feedback and that the phosphatase rate is independent of the receptor number. We also assume, as above, that the phosphatase rate  is much larger than the maximal kinase rate $p \gg k_X , k_Y$ for all choices of inputs. We obtain the steady-state solution by setting the left hand sides of the above equations to zero, which yields  
\be
Z \approx \frac{k_X}{p} X + \frac{ k_Y^0}{p} (N_{Y0} +  \frac{\delta N_{Y} Z}{K+Z}),
\label{posfbeq2}
\ee
where, for simplicity, we denote the steady-state output by $Z$. This equation can be solved for $Z$ to obtain the transfer function, $f_{\rm fb}(X,Y)$, in the presence of feedback.

A particularly interesting parameter range is the regime when $ k_Y^{0}(N_{Y0}+\delta N_Y) \gg k_X \gg k_Y^{0}N_{Y0}$. In this case, the maximal kinase activity of the $X$-pathway is much greater than the maximal kinase activity of the $Y$-pathway at low $Z$, and the opposite is true at large $Z$. Thus, the positive feedback on receptor number, $N_Y$, allows the bacteria to access information preferentially about  input $X$ at low $Z$ (i.e. at high cell density) and learn preferentially about $Y$ at high $Z$ (i.e. at low cell density). 

\begin{figure}[t]
\includegraphics[scale=0.6]{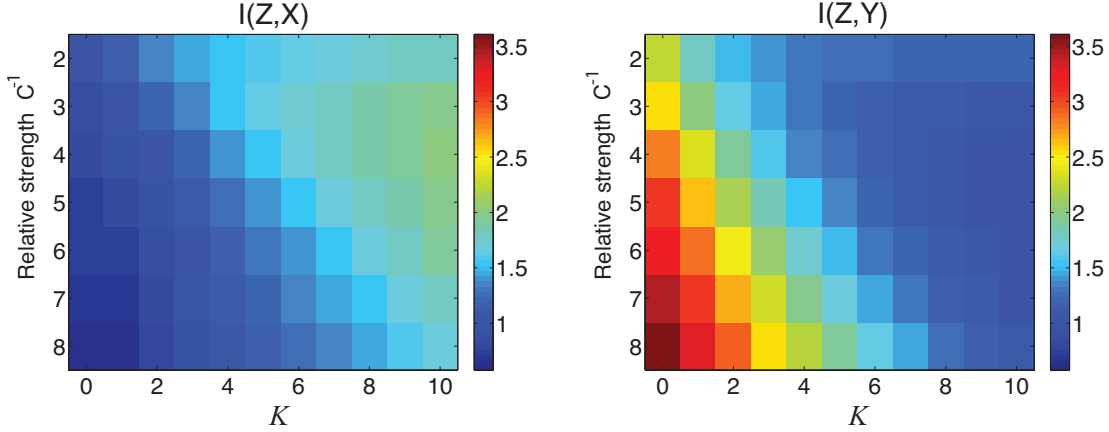}
\caption{$I(Z,X)$ and $I(Z,Y)$ for a positive-feedback architecture (Eqs. \ref{posfbeq1}, \ref{posfbeq2}) as a function of $K$ and feedback strength, $C^{-1}=(\delta C)$.}
\label{fig:SIfig}
\end{figure}

In the low-noise limit, the mutual informations $I(Z,X)$ and $I(Z,Y)$ only depend on three combinations of  parameters, the ratios of the maximal kinase activities in the presence and absence of feedback, and the half-maximal value of the feedback $K$ (data not shown). Thus, we can consider the equivalent transfer function 
\be
Z \approx X + Y(C +   \frac{ \delta C Z}{K+Z})
\ee
with $C=k_{Y}^0N_{Y0}/k_X$, $\delta C= k_{Y}^0 \delta N_Y /k_X$. We have calculated  the mutual informations $I(Z,X)$ and $I(Z,Y)$  for this transfer function using our low-noise expressions for a flat prior with inputs limited to the domain $X \ge Y$ and the results  are plotted in Fig. \ref{fig:SIfig}  for various choices of $K$ between $0$ and $10$. In order to reduce parameters we have considered the case where $C^{-1}=\delta C  = 2,3, \ldots, 8$.  Notice that by an appropriate choice of $K$, cells can learn as much, or even more, about both signals as in the absence of the feedback. This finding shows that by using a positive feedback on receptor number $N_Y$, bacteria can preferentially pay attention to AI-2 ($Y$) at low cell densities and AI-1 ($X$) at high cell densities while simultaneously learning about both input signals.

\subsection{Negative Feedback on Receptors}

We have also considered a negative feedback on receptors in the $X$-pathway. Once again the transfer function $Z=f_{\rm fb}(X,Y)$, describing the output signal (the fraction of phosphorylated output regulators), as a function of the inputs $X$ and $Y$ (the probability that the corresponding receptors are in their on states) is obtained by solving for the steady state of a set of differential equations, in this case
 \bea
 \frac{dZ}{dt} &=& (k_X X + k_Y Y) (1-Z) - p Z \nonumber \\
&=& (k_X^0 N_X  X + k_Y Y)(1-Z)-pZ,  \nonumber \\
\tau \frac{dN_X}{dt}&=&  \frac{\bar{K} N_{X0}}{\bar{K}+Z} -N_X,
 \label{negfbeq1}
 \eea
where $N_{X0}$ is the number of receptors at $X=Y=0$ and $\bar{K}$ sets the scale for the negative feedback. We obtain the steady-state solution by setting the left hand sides of the above equations to zero and recalling that $p \gg k_X, k_Y$. This analysis yields  (where for simplicity we also denote the steady-state output by $Z$)
\be
Z \approx  \frac{k_X^0N_{X0} \bar{K}}{p(\bar{K}+Z)}  X + \frac{k_Y}{p} Y.
\ee
This equation can be solved for $Z$ to obtain the transfer function in the presence of feedback, $f_{\rm fb}(X,Y)$. The mutual information is invariant under a constant rescaling  $Z \rightarrow \frac{p}{k_Y} Z$.  Thus, to reduce the number of parameters we have calculated the mutual informations $I(Z,X)$ and $I(Z,Y)$ for a family of functions (for the rescaled $Z$) of the form
\be
Z \approx  \frac{D}{K+Z}X +Y, 
\ee
with $D= \bar{K} k_X^0N_{X0}p/k_Y^2$ and $K= p \bar{K}/k_Y$, and the results are shown in Fig. \ref{fig:SIfig2}. In the plots we have used a flat prior with inputs restricted to the domain $X \ge Y$.

\begin{figure}[t]
\includegraphics[scale=0.8]{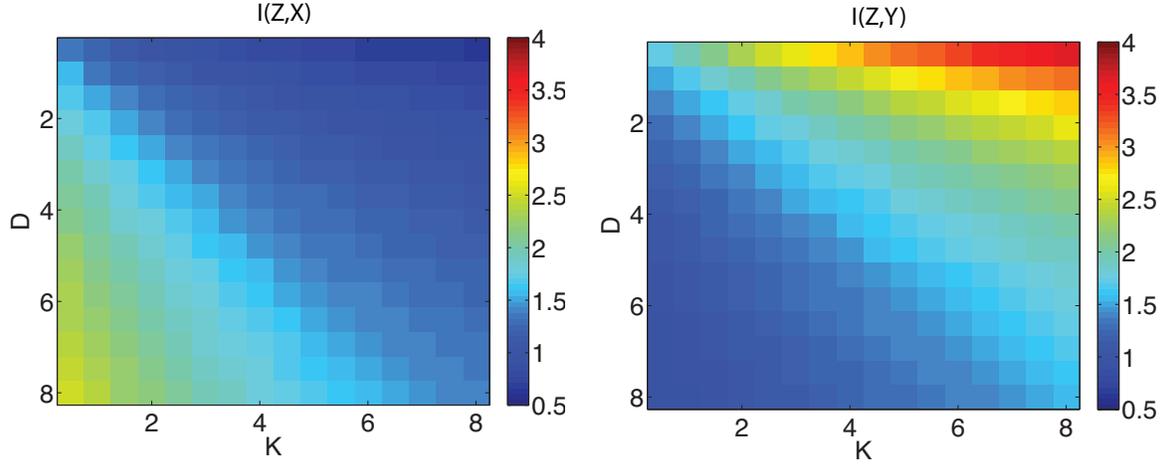}
\caption{$I(Z,X)$ and $I(Z,Y)$ for a negative feedback architecture (Eq. \ref{negfbeq1}) as a function of $K$ and $D$.}
\label{fig:SIfig2}
\end{figure}

\end{document}